# Comparative Study of Edge-Localized Mode (ELM) Heat Load Effects on Plasma-Facing Materials Using Runaway Electrons in the Damavand Tokamak

Ali Masoudi, Davoud Iraji[*], Chapar Rasouli

*Department of Physics and Energy Engineering, Amirkabir University of Technology, Tehran, Iran*

[*]E-mail: Iraji@aut.ac.ir

**Abstract**

Edge localized modes (ELMs) and runaway electrons (REs) pose significant challenges for all Tokamak devices. Both phenomena act as potent heat sources, potentially shortening the lifespan of plasma-facing materials (PFMs). These thermal loads can manifest in various detrimental effects, including melting, sputtering, cracking, blistering, and other forms of material degradation. While the ELMs are an intrinsic feature of H-mode operation in Tokamaks, runaway electrons pose a potential threat across all Tokamak device scales.

In devices such as ITER, even with mitigation strategies, the ELMs can still impose considerable heat loads on the PFMs, reaching levels of approximately 1 MJ/m².

Various methods exist to experimentally simulate the heat load effects of ELMs on PFMs. In this study, the thermal loads from runaway electrons in small-scale Tokamaks are considered for this purpose.

The presence of small-scale Tokamaks, exemplified by the Damavand experiment, facilitates the investigation of runaway electron energy deposition.

Analysis of the experimental data indicates that the REs populations generated by plasma instabilities within the Damavand Tokamak discharges exhibit heat densities on the order of MJ/m² to 1 $cm^2$ area PFMs. Furthermore, considering an average REs energy of 1 MeV and the prevailing discharge current, calculations indicate that the average total energy of the REs population per discharge is approximately 1 kJ, which is subsequently deposited on the Tokamak limiter. Considering the limiter's surface area and assuming that only 40% of the REs energy is transferred to it, the calculated heat load density reaches the order of MJ/m², comparable to that observed during the ELMs events in large-scale fusion devices.

This correspondence in heat flux levels enables the Damavand Tokamak to investigate material behavior under intense thermal loads, offering valuable insights into the effects of the ELMs on the performance and durability of PFMs in large-scale Tokamaks.

## 1. INTRODUCTION

Fusion devices are consistently under the threat of the problems arise from the impacts of thermal loads on plasma-facing materials (PFMs) [1–7]. The resulting thermal loads induce surface cracks, melting, splashing, and in addition to component damages, lead to plasma contamination and probably termination of confinement [8]. On the other hand, replacement or repairing of components face notable concerns after the operation of large-scale devices like ITER. Hence, it is imperative to undertake comprehensive research on the PFMs [9].

There are different transient heat sources that deposit great amounts of energies on materials, besides the continuous thermal loads from edge particles to plasma-facing components (PFCs) during the normal operation of Tokamaks [10].

As of now, the EAST tokamak has achieved a record-breaking plasma duration of 1056 seconds (at the electron temperature and density of 6.5 keV and $1.8\times10^{19}$ $m^{-3}$, respectively), during which the thermal peak flux to the divertor plates reached approximately 3 MW/$m^2$ [11–13].

The pulse length of the plasma discharge in ITER is planned to be 450 seconds, subjecting the PFCs to quasi-steady thermal loads of up to 20 MW/$m^2$, leading to recrystallization and fracture of connectors [14].

A distinct challenge comes from the transient loads imposed by the core plasma during the Edge Localized Modes (ELMs), which are characterized by higher density and energy compared to the edge particles [15]. The ELMs are inherent characteristics of Tokamak operation in high-confinement modes (H-Modes). The ELMs, are also known as rapid Magnetohydrodynamic (MHD) events with pulse durations typically ranging from 100 to 200 microseconds, are driven by pressure gradients and can adversely affect the PFMs [16–18]. Consequently, optimal PFCs need to be composed of materials that exhibit resistance to the ELMs [19].

Different types of the ELMs occur in Tokamaks. The ELMs with higher amplitudes tend to have lower frequencies [17]. Following the ELMs type I, recognized as the most serious form, up to 10% of the energy of core plasma is transferred to the PFCs surface [20].

The ELMs, can impose the energies up to GW/$m^2$ with pulse lengths of up to 500 µs (based on some references the duration can reach 1 ms [21]) and frequencies of several Hz to the PFCs. These loads can lead to material fractures and dust formation [12].

The overall energy release during disruptions in ITER is estimated at 120-175 MJ, and therefore, 17.5 MJ anticipated to be released during the ELMs [20]. Even at low frequencies of 1-10 Hz, this energy release is sufficient to significantly heat the PFCs (the energy limit for tungsten as the divertor material is 0.2-1 MJ) [22]. Likewise, in the HL-2A Tokamak, the ELM-induced heat flux to the divertor has been measured to be in the range of 1.5-3.2 MW/$m^2$ [23].

Controlling the ELMs and mitigating their effects through techniques such as pellet injection, resonant magnetic perturbation and active water cooling of components have been subject of ongoing research and development efforts [24,25]. However, the current approaches to ELM mitigation have not been shown to be sufficiently effective in eliminating the heat loads associated with the ELMs [26].

Generally, it is estimated that even the mitigated ELMs can apply thermal loads of up to 1 MJ/m$^2$ at frequencies of 50 Hz on the PFCs [27–29].

Moreover, the typical ELMs frequency at ITER is 25 Hz, and the flat-top phase lasts for 400 seconds, resulting in $10^4$ ELM events during each discharge [12]. Therefore, within an operational period comprising ten discharges in ITER, an estimated $10^5$ ELM events are anticipated, posing a significant risk of irreparable damage to PFCs. For this reason, the ELMs represent a critical area of investigation within fusion research.

Given the critical importance of investigating the heat load effects of ELMs on plasma-facing components, and considering the high costs of conducting related experiments in large-scale tokamaks, various experimental methods—such as using lasers, electron beams, and plasma guns—have been employed to simulate ELM heat loads. It appears that runaway electrons can also induce similar thermal effects under more accessible and cost-effective conditions using small-scale tokamaks, which is the focus of this research [30–35].

Runaway Electrons are an intrinsic feature of Tokamak plasmas, occurring across all device configurations, including both H-mode and L-mode operations. The energy deposition from REs onto plasma-facing materials can cause significant damage, particularly during disruptions, where REs capture a considerable part of the plasma's magnetic energy. In the EAST Tokamak, intense heat loads during experimental operations resulted in significant material degradation of the PFCs. This degradation manifested as substantial melting, exfoliation, and droplet ejection. Furthermore, the REs contributed to the observed damage, inducing pronounced melting of the divertor dome plates. The predicted generation of the REs beams with energies reaching 100 MJ in ITER, capable of depositing an energy flux of approximately 30 MJ/m² on the PFCs, has profound implications for the design and materials selection in confinement fusion devices. This necessitates the development of materials with exceptional resistance to thermal loads [36–40].

Plasma electrons can enter the runaway regime through several mechanisms, including the Dreicer mechanism, the hot-tail mechanism, and the avalanche process. Runaway electrons (REs) are typically generated during plasma disruptions, but they can also be produced during tokamak discharges, particularly at low plasma densities [41–44].

Plasma disruption contributes to the generation of runaway electrons (REs) through several phases. The first phase is the thermal quench, during which the rapid cooling of the plasma can quickly initiate the generation of REs [45,46]. The second phase is the current quench, which is triggered by the sharp increase in plasma resistivity that occurs during the thermal quench. The subsequent plasma current decay, occurring over a longer time scale, induces a large toroidal electric field through Faraday's law [47]. This induced field is proportional to the current decay rate. Because of the plasma current's high decay rate, the resulting electric field becomes strong enough to strongly accelerate bulk plasma electrons to high energies in the third phase of disruption. Actually, during a disruption, the enhanced force exerted by the toroidal electric field on electrons (opposing the frictional force from background ions) results in electron acceleration to relativistic speeds and their subsequent departure from the distribution function. In other words, above a critical speed, the electric field's force dominates over collisional friction, which defines the Dreicer mechanism [42–44,48,49].

During disruptions, if the induced electric field exceeds a critical threshold $E_D$ ($E_D = \frac{n_e e^3 \ln\Lambda}{4\pi\varepsilon_0^2 T_e}$), the Dreicer mechanism is triggered [48].

Runaway electrons can also be produced by other key mechanisms. The hot tail mechanism represents another pathway for runaway electron production in the first phase of disruptions, before the induction of a high electric filed. When the thermal quench duration is shorter than the collision time for electrons near the runaway threshold energy in a quasi-steady state, energetic electrons cannot cool and equilibrate with the Maxwellian bulk, instead forming a high-energy tail that enhances the runaway region population [41–44].

Although, the generation of runaway electrons is significantly suppressed below the critical electric fields, the number of REs can gradually increase over time through Coulomb collisions between the bulk plasma electrons [50,51].

Another mechanism active during plasma discharges is the avalanche process, which occurs notably at low plasma densities. The Avalanche mechanism for runaway electron generation involves a process where existing energetic electrons, undergo knock-on collisions with thermal electrons. When the electric field exceeds the critical field $E_C$ ($E_C = \frac{T_e E_D}{m_e c^2}$), these collisions can transfer sufficient energy to kick thermal electrons into the runaway regime, leading to multiplication of runaway electrons and creating a cascading effect [50,51].

In the interaction of runaway electrons with the PFCs within Tokamaks, Parail-Pogutse instabilities serve as a key factor. These instabilities, driven by the anomalous Doppler effect, appear periodically and cause a rise in the transverse energy of the electrons. Their appearance in tokamak discharges is considered an indicator of the presence of runaway electrons. Runaway dominated (R.E.D) discharges are also marked by an absence or minimal presence of Mirnov oscillations. The emission of hard X-rays bursts is another indication of these discharges [52].

All in all, the REs are those electrons that finally overcome the magnetic confinement and imping upon the PFCs. When the REs interact with plasma-facing components, they deposit their energy through several primary mechanisms, such as ionization, Bremsstrahlung radiation, Compton scattering, the photoelectric, and multiple scattering processes. These interactions may also give rise to other products. Among them, secondary electrons which may have enough energy to escape the Tokamak vessel. Meanwhile, the production of secondary neutrons, a consequence of gamma-neutron activation of the PFCs following Bremsstrahlung radiation induced by the interaction of the REs with PFCs, constitutes an adverse effect [53].

Reports from Tokamak experiments indicate that runaway electrons can carry a significant fraction of the total plasma current. These energetic particles, capable of attaining energies on the order of MeV, pose a severe threat to the integrity of the PFCs [54,55].

Additionally, small Tokamaks have been shown to support discharges in which runaway electrons generate the entirety of the plasma current. Techniques applied in the Damavand Tokamak have proven the feasibility of creating R.E.D regimes, where nearly all plasma current is transformed into runaway current [52].

This work aims to reproduce the impact of ELM heat loads on plasma-facing components by using the Damavand Tokamak, operated in a fully runaway electron regime, as an accessible and cost-effective experimental platform. Within this framework, the energy, interaction time, and penetration depth of REs on PFCs are quantified, and representative evidence of the resulting damage is presented. The central goal is to establish that the thermal effects induced by REs in small-scale Tokamaks can serve as a reliable analogue to the energy impact of ELMs in large-scale devices, thereby demonstrating that the two events are fundamentally comparable.

This paper is organized as follows: The experimental setup which is used to form, measure and study the behavior of the REs is described in Section 2. Section 3 explains the analytical methods and discusses the experimental results, including the calculation of the energy of the REs responsible for generating hard X-rays during a discharge. It also estimates the total energy of the REs released in the plasma discharge and the energy density transferred to the Tokamak limiter. This energy is compared to the energy of the ELMs in large-scale Tokamaks. Eventually, the conclusions of this study are presented in Section 4.

## 2. Experimental Setup

The Damavand Tokamak is a small size device characterized by a major radius of 36 cm and a plasma cross-section with parameters b = 10 cm and a = 7 cm. It achieves a maximum toroidal magnetic field of 1.2 T and operates with plasma currents ranging from $I_P$ = 28 kA to 38 kA, with peak current reaching $I_P$ = 40 kA. The maximum ion temperature is approximately $T_i$ = 150 eV, while the maximum electron temperature is $T_e$ = 300 eV. The plasma discharges last for approximately 22 ms [30,56]. Two limiter slabs have been installed on the low-field side (LFS) and high-field side (HFS) of the Tokamak. Hard X-rays (HXRs) with energies of about 200 keV up to 3 MeV, emitted during discharges, are detected using NaI scintillator detectors positioned approximately 3 meters from the periphery of the Tokamak vessel. These high energy photons indicate the generation of the REs [57]. Poloidal Magnetic field measurements in this Tokamak are conducted using 48 pickup coils installed within two stainless steel ducts at various locations on the HFS and LFS around the vacuum vessel. The signals from these probes are integrated using distinct time constants, and the amplified pulses are captured and recorded. During each discharge of the Damavand Tokamak, a data acquisition (DAQ) system acquires loop voltage, plasma current, HXR signals, and the signals received from magnetic probes simultaneously, and with the sample rate of 2 Msample/ (s channel) [50].

## 3. Results and discussion

As stated before, the Damavand Tokamak is an appropriate devise, able to operate in the R.E.D regimes. The selected method in this study to reach this goal has been the reduction of the pre-ionization energy [52].

In R.E.D regimes, runaway electrons acquire energy from the toroidal field during each rotation, leading to extremely high energy levels. Assuming that each RE does not traverse the same point on the torus more than once, the number of runaway electrons can be expressed as:

$$n = \Delta q/e \quad (1)$$

That Δq refers to the electric charge of the REs, and it is calculated as:

$$\Delta q = I\Delta t \quad (2)$$

To calculate the actual number of the REs, the number n should be divided by the number of turns the REs make in the toroidal electric field. The number of turns (k) is determined by dividing the distance between their formation point and stopping point by the length of the Tokamak torus:

$$k = \frac{\Delta x}{2\pi R_0} = \frac{\bar{v}\Delta t}{2\pi R_0} \quad (3)$$

$R_0$ represents the main radius of the Tokamak, Δx is the transverse distance of the REs, Δt is the lifetime of the REs, and $\bar{v}$ is the average velocity of the REs, which can be calculated by knowing their average kinetic energy [58]:

$$\bar{E} = \gamma m_e c^2 = \frac{m_e c^2}{\sqrt{1 - \frac{\bar{v}^2}{c^2}}} \quad (4)$$

In this equation, γ is the Lorentz factor, $m_e$ refers to the mass of the electron, and c represents the speed of light [58].

The average kinetic energy of the REs in the Damavand Tokamak is approximately 1 MeV, which results in an average velocity of 0.86c, as derived from Eq. (4) [52]. Therefore, for the REs in the Damavand Tokamak, the number of turns (k) is calculated using Eq. (3) and is approximately $1.34 \times 10^4$. The real number of REs is then given by:

$$n_r = \frac{n}{1.34 \times 10^4} = \frac{\Delta q}{1.34 \times 10^4 e} \quad (5)$$

Here, Δq is substituted from Eq. (2), and the calculation is carried out using the REs current (I), and their lifetime (Δt).

Runaway electrons that make up a fraction of the plasma current are energetic enough to traverse the magnetic field lines and collide with the Tokamak's vacuum vessel. Upon interacting with plasma-facing components, like the Tokamak limiters, they emit hard X-rays (HXRs). These HXRs carry enough energy to pass through the Tokamak vessel and reach the NaI detectors [50].

Based on the measurements of the emitted energy spectra of the HXRs in the Damavand Tokamak in different positions of detectors, the energy range of hard x-rays is from 0.2 MeV to 3 MeV, far exceeding the energy of the bulk plasma, which is around 300 eV for electrons and 150 eV for ions. This makes HXRs a primary indicator of the presence of the REs and their interactions with the PFCs [50]. Additionally, during runaway-dominated discharges, Tokamaks show little to no Mirnov oscillations on magnetic probes, while noticeable spikes appear in the hard X-ray and loop voltage signals [52].

The Conditional Average Sampling (CAS) method was applied to analyze data from the Damavand Tokamak, enabling the simultaneous investigation of the time evolution of toroidal current, loop voltage, and hard X-ray intensity. In this method, the HXR signal is used as the reference pulse.

Figure 1 shows the signals of discharge current, loop voltage and hard x-ray for a typical discharge of the Damavand Tokamak. In order to study the temporal behavior of runaway electrons related phenomena, the CAS method is performed for three distinct phases of ramp-up (0 to 5 ms), flat-top (5 ms to 15 ms), and ramp-down (15 ms to 22 ms).

The Current Filament Model (CFM) is another technique employed to analyze the spatial behavior of the REs during moments of instabilities in the Damavand Tokamak. This method approximates the plasma column as a finite number of current filaments. Consequently, the vertical (Z) and radial (R) positions of the plasma current were reconstructed, and a diagram of the reconstructed discharge current was derived from the CFM [52].

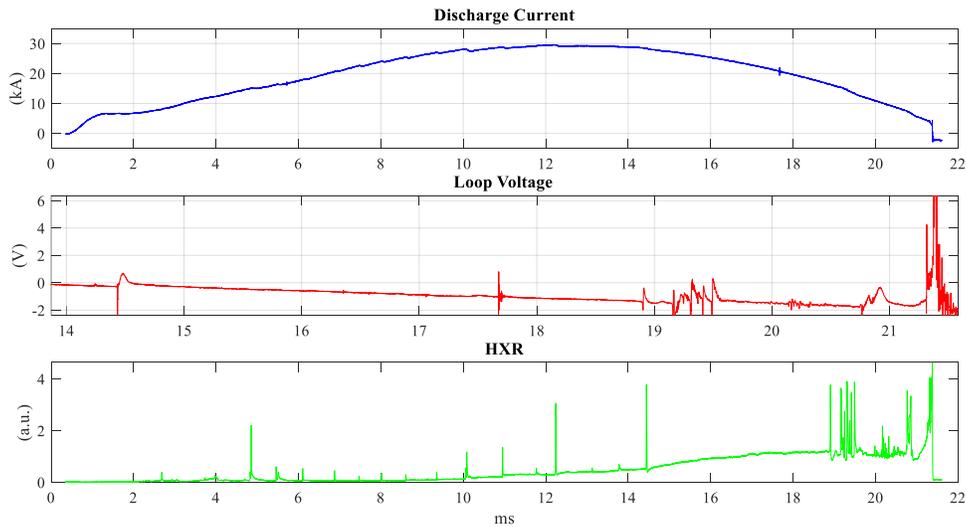

Figure 1 A plasma discharge at 1.1 T in the Damavand Tokamak with HXRs spikes (shot no. 148).

Figure 2 illustrates the CASed plasma current, loop voltage, HXR bursts, reconstructed R, reconstructed current, and reconstructed Z, recorded from 200 µs prior to the instability to 200 µs following it, during the flat-top phase of a discharge in the Damavand Tokamak. The discharge exhibits a peak plasma current of 33 kA at a magnetic field strength of 0.8 T. Data obtained using the CAS method for 26 events reveals that the instability reduces the discharge current by 125 A over 103 µs. During such instabilities, the REs beam shifts toward the HFS, generating an HXR burst as the REs collide with the Tokamak limiter. The observed reduction in plasma current and simultaneous induction of loop voltage validate the collision and disappearance of the REs beam. Therefore, the reduction in plasma current directly corresponds to the current of the REs responsible for the detected HXRs.

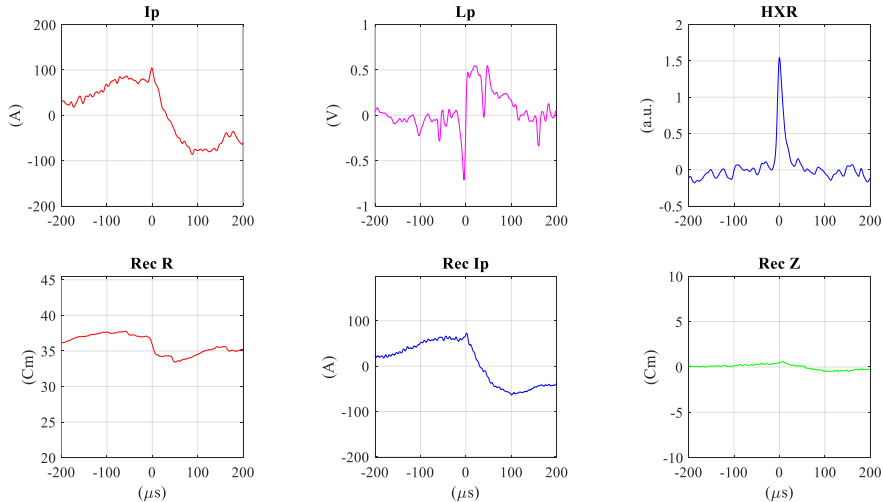

Figure 2 Profiles of toroidal current $I_p$, loop voltage $L_p$, HXR intensity, Rec R, Rec $I_p$, and Rec Z, during the flat-top phase of the discharge no. 139, when the toroidal magnetic field is 0.8 T.

The influence of the toroidal magnetic field on the position and current of the REs was examined by recording data from a plasma discharge at a magnetic field of 1.1 T. Results indicated that increasing the toroidal magnetic field reduced the number of instabilities in the flat-top phase to eight, while the peak plasma current was around 30 kA. Figure 3 illustrates the outcomes derived from applying the CAS method to these eight events [50]. From the figure, it can be deduced that the reduction in plasma current (along with the corresponding increase in loop voltage) at the moment of instability is nearly twice as large as that observed at a lower magnetic field. Specifically, the current change is approximately 370 A and occurs within about 113 µs, which is half the REs' lifetime at 0.8 T. The larger toroidal magnetic field results in fewer instabilities, larger RE currents, and shorter collision times between the REs and the PFMs, as evident from the narrower HXR peak widths.

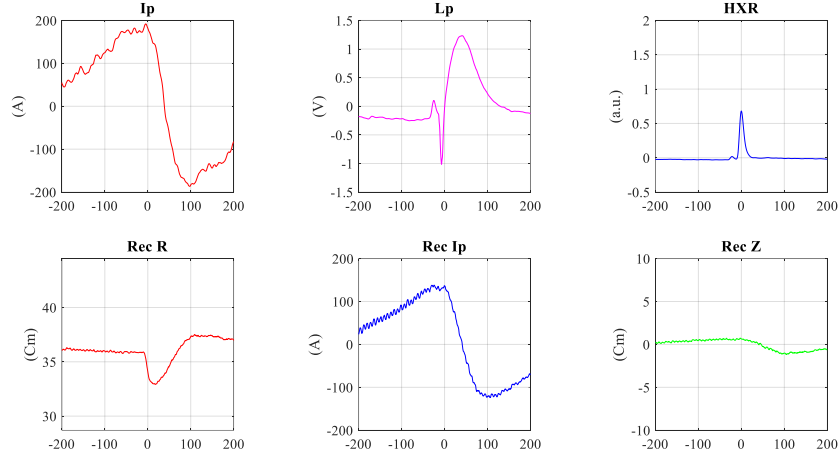

Figure 3 Profiles of toroidal current $I_p$, loop voltage $L_p$, HXR intensity, Rec R, Rec $I_p$, and Rec Z, during the flat-top phase of the discharge no. 148, when the toroidal magnetic field is 1.1 T.

The number of runaway electrons responsible for the detected HXRs is determined using Eq.(5), based on their current and lifetime. By referencing Figure 4, the mean kinetic energy of REs is found for each discharge phase ($\overline{E} = 1.175$ MeV for ramp-up, $\overline{E} = 1.35$ MeV for flat-top, and $\overline{E} = 0.7$ MeV for ramp-down) at a magnetic field of 1.1 T. This allows for the calculation of the total energy of the REs corresponding to each HXR burst (single event):

$$E_{\text{single event}} = n_r \overline{E} \tag{6}$$

The entire energy of the REs contributing to the detected X-ray photons is obtained by multiplying the energy per event by the number of instability occurrences included in the CAS evaluation:

$$E_{\text{REs}} = n_x n_r \overline{E} \tag{7}$$

Here, $n_x$ represents the number of the HXRs bursts.

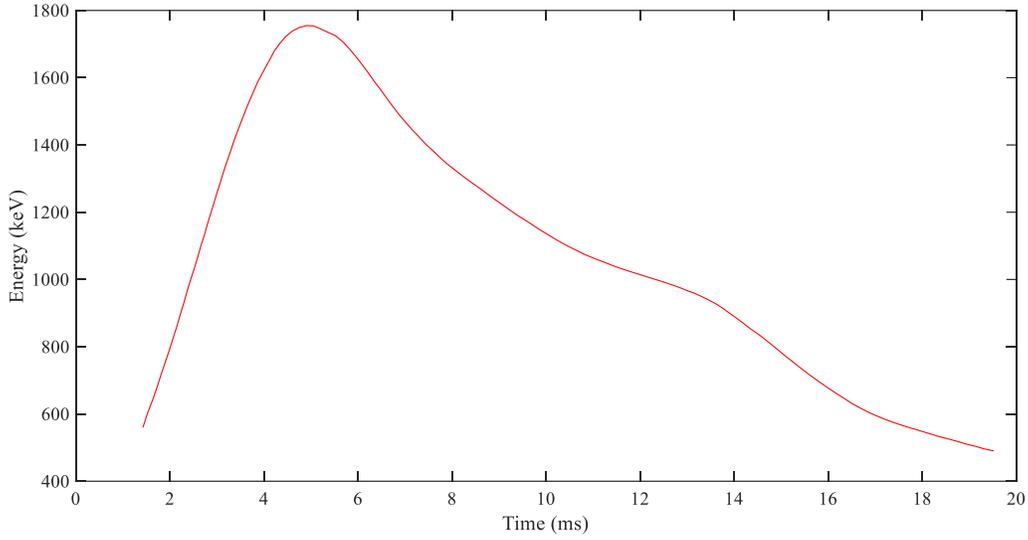

Figure 4 The temporal progression of the average kinetic energy of the REs throughout the plasma discharge in the Damavand Tokamak [51].

During the flat-top phase, runaway electrons with a current of 370 A and a lifetime of 113 µs, confined in a 1.1 T toroidal field, have an electric charge of 0.037 coulombs per event (as determined from Eq. (2)). Accordingly, the actual number of runaway electrons ($n_r$) is calculated using Eq. (5) to be approximately $1.9 \times 10^{13}$, and the total kinetic energy for a single event in the flat-top phase is about 3 J. Since 8 events were CASed to produce the diagrams for the flat-top phase within this magnetic field, the total energy of the runaway electrons ($E_{REs}$) is about 24 J.

To calculate the total energy of all runaway electrons in a discharge, it is essential to account for their energy contributions during the ramp-up, flat-top, and ramp-down phases. Figure 5, generated using the CAS method for 6 events, provides an analysis of the ramp-up phase at 1.1 T. The results indicate that the current and lifetime of the runaway electrons are 124 A and 205 µs, respectively. The real number of runaway electrons is approximately $1.2 \times 10^{13}$, and the energy of the detected HXRs, produced by the interaction of the runaway electrons with the limiter, is about 12 J.

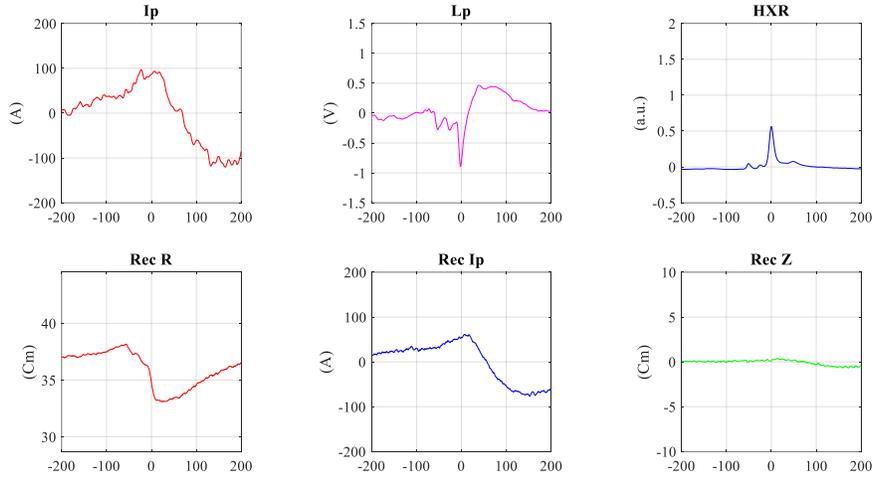

Figure 5 Profiles of toroidal current Ip, loop voltage Lp, HXR intensity, Rec R, Rec Ip, and Rec Z, during the ramp-up phase of the discharge no. 148, when the toroidal magnetic field is 1.1 T.

Figure 6 shows the results of using the CAS technique on 14 events to examine the ramp-down phase at 1.1 T. The figure indicates that the current of the runaway electrons is approximately 81 A, with a lifetime of 63 µs. Therefore, the number and energy of the runaway electrons during the ramp-down phase in this toroidal field are estimated to be $2.4 \times 10^{12}$ and 5 J, respectively.

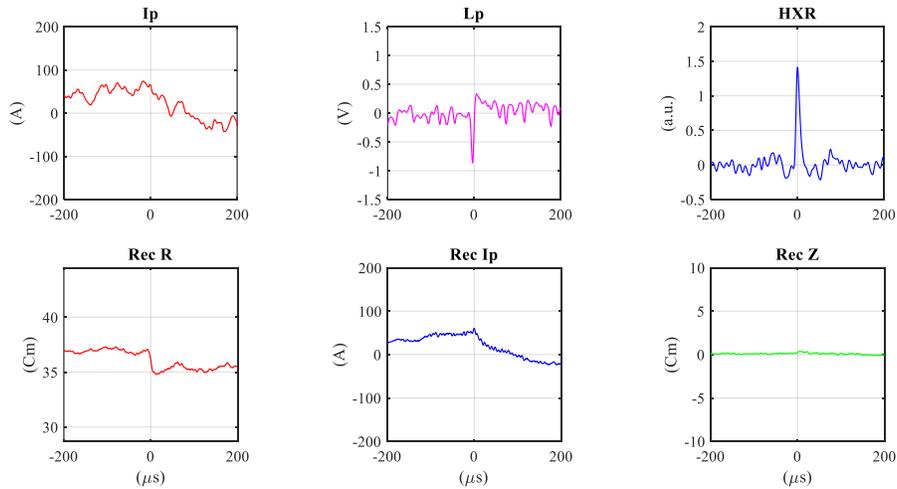

Figure 6 Profiles of toroidal current $I_p$, loop voltage $L_p$, HXR intensity, Rec R, Rec $I_p$, and Rec Z, during the ramp-down phase of the discharge no. 148, when the toroidal magnetic field is 1.1 T.

Apart from the instabilities discussed, a significant drop in the plasma current is commonly observed, which corresponds to the termination stage of the plasma discharge (as it is seen in Figure 1), resulting in the generation of an intense HXR peak. Because of the limits the HXRs detection system, distinct bursts of high intensity emissions cannot be detected. Therefore, these events are not considered in the CAS analysis. The current drop at the ending phase of the plasma represents the collision of the REs with PFCs. At this stage, for a magnetic field of 1.1 T, the REs current and lifetime are approximately 7 kA and 238 µs, respectively. Accordingly, the estimated number of runaway electrons at the plasma termination ($\bar{E}$ = 0.45 MeV) is $77.5 \times 10^{13}$, with a total energy of 56 J.

To summarize, the total energy generated by the REs during each discharge of the Damavand Tokamak amounts to at least 97 J. This energy is associated with REs that reach the Tokamak limiter, producing bremsstrahlung radiation detectable by scintillators. When a sample with a 1 cm length is inserted into the plasma as a part of the limiter, and considering a plasma section with a 7 cm diameter, 1/7 of the plasma flux is directed onto the sample surface. As a result, the minimum heat load on the sample from the REs is 14 J. This thermal load corresponds to a heat load density of 0.13 MJ/m$^2$ per shot when impacting the surface of a PFC, which is comparable to the heat load caused by ELM events on PFMs. In practice, the heat load density from five shots in the Damavand Tokamak during the R.E.D. regime is equivalent to that experienced from ELMs. Moreover, the HXR bursts shown in the figures indicate that the duration of RE–PFM interactions in this tokamak is on the order of 100 µs—specifically, 113 µs in flat-top, 63 µs in ramp-down, 205 µs in ramp-up, and 238 µs at the end of the discharge—comparable to typical ELMs duration, which is of the order of hundreds of µs.

A summary of the calculations is presented in Table 1.

It is also important to note that, after prolonged operation of Tokamaks, the frequent occurrence of heat loadings can lead to fatigue failure of the PFCs [59].

Table 1 Energy of runaway electrons in Damavand Tokamak.

| Discharge Phase | Ramp-Up | Flat-Top | Ramp-Down | Termination Stage |
|---|---|---|---|---|
| **Energy from REs per Phase (J)** | 12 | 24 | 5 | 56 |
| **Total Energy of REs in each Discharge (J)** | | 97 | | |

The total energy of the REs in the Damavand Tokamak can also be estimated through an alternative scenario as follows. During each plasma discharge, the total electrical energy of the REs is:

$$E_{elec}= \int V(t)I(t)dt \approx \overline{V}_{loop}\overline{I}\Delta t \qquad (8)$$

Here, V(t) and I(t) denote the loop voltage and plasma current at time t, respectively, while, $\overline{V}_{loop}$ and $\overline{I}$ indicate the average loop voltage and average plasma current, and $\Delta t$ is the duration of the discharge.

Assuming an average loop voltage of 7 V, an average plasma current of 15 kA, and a shot duration of 20 ms, the total electrical energy of the plasma in a typical discharge of the Damavand Tokamak is estimated to be around 2.1 kJ. Therefore, in R.E.D. discharges, the total energy of the runaway electrons reaching the Tokamak limiter is generally in the order of kilojoules. Reports indicate that at least 40% of this energy is transferred when the runaway electrons impact the surface of the Tokamak limiters. With each limiter having an area of 7 cm × 1 cm, a heat load density on the order of 0.3 MJ/m$^2$ is transferred to the limiter during each Tokamak shot [41].

At this point, the heat loads of the REs in small-scale tokamaks and those of ELMs, regarding their effects on the PFCs, will be compared:

1) The energy deposition from the REs during few shots of Damavand Tokamak was calculated to be at the order of MJ/m$^2$, which is comparable to that of the ELMs in large-scale Tokamaks [20,25,27–29].
2) Temporal evolution of the REs in this Tokamak was stated to be on the order of a hundred microseconds, which is similar to the duration of ELM bursts [16–19].
3) The energy deposition depth of the REs has been calculated in Ref. [41], to be on the order of hundreds of micrometers for the energy range of REs in the Damavand Tokamak (with an average energy of about 1 MeV). This depth is again comparable to the energy deposition depth of ELMs in large-scale Tokamaks.

Moreover, as detailed in Ref. [60], the effects of ELM-induced heat load on tungsten PFCs in the JET Tokamak are depicted in Figure 6 of that study. The image captures a distinct melted edge on the tungsten surface, along with evident layering of material and a ~150 μm droplet adhered to the sidewall. The droplet's layered morphology and associated cracking are also clearly visible.

Additionally, results from numerous experimental simulations investigating the effects of ELM-induced heat loads on PFCs have been reported in various studies ([19,22,35,61–67]), clearly demonstrating several types of material degradation, such as cracking. For instance, Figure 6 from Ref. [65] illustrates the development and extent of cracks in pure tungsten, W–1 vol% $Y_2O_3$, and W–5 vol% $Y_2O_3$, after being subjected to laser-induced heat loads of 0.38 MJ/m² and 1.13 MJ/m², which correspond to crack depths of 300–400 µm and 850–1500 µm, respectively. Similarly, Figures 3 and 5 in Ref. [64] depict typical damage features—including cracking and surface swelling—on PFCs following exposure to ELM-like heat loads induced by laser pulses.

Further observations are provided in Figure 3 of Ref. [63], where surface roughening, melting, and the formation of shale-like microstructures on tungsten samples are evident after laser pulses at an energy density of 0.76 MJ/m². Likewise, Figure 6 from Ref. [61] shows crack formation on a tungsten PFC subjected to a laser heat load of 0.1 MJ/m².

In experiments simulating ELM effects using electron beams, Figure 1 of Ref. [22] presents evidence of crack formation and the presence of a molten layer on a PFC at the heat load of 0.735 MJ/m², performed at the JUDITH facility.

Additionally, Figure 3 from Ref. [35] demonstrates extensive melting, large-scale melt flow, and typical surface cracking within the molten layer of a tungsten sample after it was subjected to five electrothermal plasma pulses of 1.25 MJ/m² at a frequency of 1 Hz. Under single-pulse conditions at the same heat load density, the depth of the melting layer reached approximately 100 µm.

These types of damage phenomena—including cracking, surface melting, and microstructural degradation—are notably similar to the damage effects induced by REs in several tokamaks [60].

Experimental observations from Damavand reveal severe limiter degradation resulting from RE-induced thermal loads on plasma-facing components. As demonstrated in Figure 7, the surface exhibits substantial melting, cracking, and critical material degradation. Figure 7 a and Figure 7 b clearly illustrate melting at the limiter edge and surface melting, respectively. The movement of the molten layer is evident in Figure 7 d, with Figure 7 e revealing the associated surface cracking. Figure 7 c and Figure 7 f display surface blisters formed as a result of thermal loading by the REs.

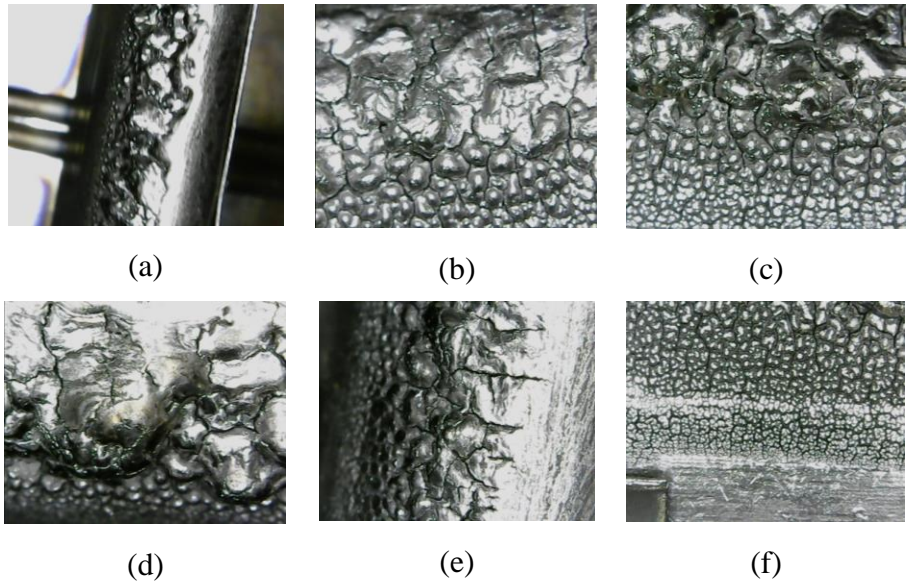

| (a) | (b) | (c) |
| (d) | (e) | (f) |

Figure 7 Effects of runaway electron (RE) thermal loads on the stainless-steel limiter of Damavand Tokamak, showing melting, cracking, and blistering.

Finally, it can be inferred once again that the heat load of the REs on the PFCs in the Damavand Tokamak, is comparable to the energy released during ELMs events in large-scale devices such as ITER. This characteristic makes the Damavand and other small-size Tokamaks intriguing test beds for investigating the adverse effects of the heat loads from ELMs events on the PFCs [12,18,29,68].

Table 2 shows some characteristics of the REs in the Damavand Tokamak and the ELMs.

Table 2 Characteristics of REs in Damavand and the ELMs.

| | Heat Load (MJ/m$^2$) | Energy Deposition Depth (μm) | Duration (μs) | Typical Damage Type | Reference |
|---|---|---|---|---|---|
| **REs in Damavand** | 1 (~ 5 shots) | 400 | 60-250 | Cracking, Melting, Blistering | [41,60,69] |
| **ELMs** | 1 | 350 | 100-200 | Cracking, Melting, Erosion, Droplet Formation | [19,22,35,60–67] |

## 4. Conclusions

The study of the PFCs has always been of paramount importance in the field of fusion energy research. Fusion devices must be engineered to withstand the detrimental effects arising from the degradation of these components. Consequently, this work was undertaken with the primary objective of investigating the impact of heat loads on the performance and longevity of the PFCs.

As the ELMs are among the most significant heat sources in Tokamaks that adversely impact the PFCs, this study focuses on the heat loads associated with these events. To this end, the investigation encompasses the kinetic energy of the REs, another critical heat source in Tokamaks, and their subsequent heat transfer to the PFCs within the Damavand Tokamak.

The Damavand Tokamak, configured for operation within the R.E.D regimes, presents a valuable platform for investigating these phenomena. Prior researches conducted on the Damavand Tokamak have established the mean energy of 1 MeV per runaway electron. The present analysis investigates the energy of the REs produced during each discharge of the Tokamak using the CAS method, through which diagrams of plasma current, loop voltage, and HXR spectra are obtained. With the RE energy already provided in Table 1, and the study indicates that the average total energy density deposited by these energetic particles upon the PFCs with the surface area of 1 cm² reaches the order of 0.1 MJ/m² during each discharge.

Furthermore, an alternative calculation, assuming a fully runaway electron population, predicts that an estimated 1 kJ of energy will impact the Tokamak limiter during each discharge. It is anticipated that at least 40% of this energy will be transferred to the limiter surface, resulting in a calculated heat load density on the limiter of 0.3 MJ/m$^2$.

Therefore, it is concluded that the heat load from the REs on the Damavand Tokamak limiter is in the range of 0.1-0.3 MJ/m$^2$ per shot. Moreover, reports indicate that mitigated ELMs in the ITER Tokamak are expected to subject the PFCs to heat loads of approximately 1 MJ/m$^2$.

This analysis demonstrates that the observed energy release from runaway electrons in the Damavand Tokamak exhibits a magnitude comparable to the energy transfer associated with the ELMs in large-scale Tokamak devices. On the other hand, at the average kinetic energy of the REs in small-size Tokamaks the penetration depth of heat loading is comparable to that of the ELMs in large-scale Tokamaks. Therefore, the REs in small-size devices cause surface damages of the PFCs, similar the ELMs events in large-scale devices.

The presented findings demonstrate that the Damavand Tokamak, a relatively compact and low-cost device capable of operating in the R.E.D regimes, can generate significant heat loads on the PFCs, comparable to those anticipated in larger, more complex and expensive Tokamaks during the ELMs. This research highlights the invaluable contribution of small-scale Tokamak platforms in advancing the understanding of material behavior and component survivability under the extreme thermal fluxes encountered in future fusion reactor environments. This has facilitated the study of the effects of ELMs on various plasma-facing materials.